\documentclass[12pt]{article}

\usepackage{todonotes}
\usepackage{amssymb}
\usepackage{makeidx}
\usepackage{epsf}
\usepackage{graphicx}       

\makeindex

\def\fnote#1#2{\begingroup\def\thefootnote{#1}\footnote{#2}\addtocounter{footnote}{-1}\endgroup}

\def\a{{\alpha}}       \def\g{{\gamma}}  
\def\e{{\epsilon}}     
\def\si{{\sigma}}       

 \def\G{{\Gamma}}        

\def\cA{{\cal A}}     
  \def\cH{{\cal H}} 
  \def\cL{{\cal L}}

 \def\phidot{{\dot{\phi}}}  \def\chidot{{\dot{\chi}}}   \def\rhodot{{\dot{\rho}}}  
 
    \def\rmGeV{{\rm GeV}}

   \def\rmPl{{\rm Pl}}

               \def\rmSL{{\rm SL}}

\def\rmconst{{\rm const}}

        \def\rmint{{\rm int}}

     \def\rmkin{{\rm kin}}

         \def\rmosc{{\rm osc}}

\def\rmrh{{\rm rh}}

          \def\rmtri{{\rm tri}}

 \def\mathR{{\mathbb R}}
\def\mathZ{{\mathbb Z}}

\def\beq{\begin{equation}}
\def\eeq{\end{equation}}
\def\bea{\begin{eqnarray}}
\def\eea{\end{eqnarray}}

\let\nn=\nonumber
\def\notin{\ \hbox{{$\in$}\kern-.51em\hbox{/}}}

\def\ra{{\rightarrow}}

\def\del{{\partial}} 

\def\notdiv{{\relax{~|\kern-.34em /~}}}

     \def\opsi{{\overline{\psi}}}

\begin{document}

\phantom{\hfill \today}

\vskip 1truein
\parskip=0.15truein
\baselineskip=19pt

\centerline{\Large {\bf Multifield Reheating after Modular $j$-Inflation }}

\vskip .2truein

\centerline{\sc Rolf Schimmrigk\fnote{1}{
  netahu@yahoo.com; rschimmr@iusb.edu}}

\vskip .2truein

\centerline{Dept. of Physics}

\centerline{Indiana University South Bend}

\centerline{Mishawaka Ave., South Bend, IN 46634}

\vskip 1truein

\begin{quote}
  \centerline{\bf Abstract.} 
  
In the inflationary framework of cosmology the initial  phase of rapid expansion has to be followed by a reheating 
stage, which is envisioned to end in a radiation dominated big bang. Key parameters that characterize this 
big bang state are the temperature at the end of the reheating stage and the baryon asymmetry. 
For general interacting theories these parameters are difficult to obtain analytically because of the involved 
structure of the potential. 
In this paper multifield reheating is considered for interacting theories in which the inflaton trajectory 
is weakly curved.  This scenario is realized in the model of $j$-inflation, a particular example of modular inflation, 
allowing an estimate of the reheat temperature.
\end{quote}

\renewcommand\thepage{}
\newpage
\pagenumbering{arabic}

\vfill \eject

\baselineskip=12pt
\parskip=0.02truein

\tableofcontents

\vskip .7truein

\baselineskip=22pt
\parskip=0.12truein
\parindent=0pt

\section{Introduction}

In order to understand the physical significance of the CMB observations made over the past two decades in the context of 
an inflationary picture, it is important to think in terms of a multifield theoretic framework  (a recent review can be found in  \cite{g16}). 
While few-parameter fits in the  context of singlefield inflation are compatible with the data, they provide an incomplete view
of the impact of the experimental results on the restrictions of the theory space.
An example of such a structured space  is given by the framework of automorphic inflation, an approach to 
the early universe based on symmetry principles, motivated in part by the idea to constrain the potentials with an arbitrary
 number of fields by retaining a discrete symmetry of the theory as the result of 
a weakly broken continuous symmetry \cite{rs14, rs15, rs16}.  This provides a natural embedding of the shift symmetry
into a canonical group theoretic framework.

The resulting discrete group $\G$ and the potential $V_\G$ associated to $\G$ are specified by 
numerical characteristics that allow the identification of distinct regions in the curved field theory space of the 
inflationary models, described by the field space metric $G_{IJ}$ and $V_\G$.  The restriction of automorphic 
field theory to two-field theories leads to the class of modular inflation models with a nontrivial geometry given 
by the hyperbolic Poincar\'e metric. 
Subsequent related work on modular inflation was done in \cite{bl17}, while work with a 
focus on the hyperbolic target space of modular inflation includes \cite{b17, mm17}.

After the end of inflation the reheating stage is envisioned to provide the evolutionary 
 link between the inflationary phase and the start of the radiation dominated big bang. 
 Reheating has been discussed mostly in singlefield inflation, following the first papers 
 \cite{a82etal, afw82,dl82}, with a later focus on the early stages of the reheating process,
  explored first in the references \cite{tb90, dk90, kls94, stb94, kls97}.
 In the context of multifield theories the analysis becomes complicated and most work has been restricted to
  separable potentials \cite{bk08, cgj08, bbg09, bkb10, mt13, eom14, h17etal}, but interacting fields have been considered 
 for flat target spaces in refs. \cite{ l12etal, hc13, l13etal}, and a discussion admitting curved inflaton targets can be found 
 in ref. \cite{ww15, d15etal}.
 
 The most important parameter that results from the 
reheating period is the reheat temperature $T_\rmrh$. To follow the temperature analytically through the whole period of reheating,
including preheating, has turned out to be intricate, but the complications of preheating have been recognized to 
be less essential if the interest is in the temperature at the end of reheating. The observation of the present paper is that in
 interacting theories with intricate potentials it can happen that  the inflationary trajectories are weakly curved,
resulting in potentials associated to the inflationary orbit that can be approximated by a simpler singlefield potential. 
This extends the range of potentials to which the canonical methods can be applied to analyze the reheating stage.
The phenomenon of weakly curved trajectories is realized in the model of $j$-inflation, which is defined in terms 
of  a nontrivially interacting inflaton doublet.  The simplification obtained in this case
 allows to compute the reheat temperature and hence the associated baryon asymmetry.

In outline, after describing the general set-up of multifield reheating in  Section 2,  the paper introduces the general picture of weakly 
curved trajectories in Section 3. The application to $j$-inflation is given in Section 4 and the paper concludes in Section 5.

\section{Multifield reheating set-up}

 In the framework of reheating after multifield inflation it is standard to 
consider decays of various types, such as $ \phi^I~\ra~\chi^A\chi^A$ or $\phi^I\phi^I~\ra~\chi^A \chi^A$,
 parametrized by couplings of triscalar and bi-quadratic type. The Lagrangian decomposes as
    \beq
    \cL ~=~ \cL_\rmkin ~+~ \cL_\rmint,
   \eeq
   where the kinetic energy is determined by, in general nontrivial, metrics $G_{IJ}$ and $G_{AB}$
  \beq
    \cL_\rmkin ~=~ -\frac{1}{2}G_{IJ} g^{\mu\nu} \del_\mu \phi^I \del_\nu \phi^J 
       - \frac{1}{2} G_{AB} g^{\mu \nu} \del_\mu \chi^A \del_\nu \chi^B
  \eeq
and the interaction includes self-interactions as
  \beq
    \cL_\rmint ~=~  - V(\phi^I)  - U(\chi^A)  - W(\phi^I,\chi^A).
  \eeq
  The  triscalar and bi-quadratic potentials describing the interactions between the 
  inflaton and the field $\chi$ providing the decay channels can be generalized to the multifield case as
  \beq
   W(\phi^I,\chi^A ) ~=~  \sum_{I,A} \frac{1}{2} g_{IA} \phi^I (\chi^A)^2
                        +\frac{1}{2} \sum_{I,A}  h_{IA} (\phi^I)^2 (\chi^A)^2,
  \eeq
  where the triscalar coupling parameters $g_{IA}$ have dimensions of mass and the 
  bi-quadratic coupling parameters $h_{IA}$ are dimensionless. The oscillatory inflaton is obtained by
  assuming that the decay field is small at the beginning of the reheating period. 
  The dynamical equations of the inflaton are given by 
  \beq
   D_t \phidot^I ~+~ 3H\phidot^I ~+~ G^{IJ}(V+W)_{,J} ~=~ 0,
  \eeq
  where $D_t$ is the covariant derivative  and 
   the $\chi^A$ dependence in the interaction $W$ is via its correlators. 
  The decay products satisfy the equation
  \beq
 \ddot{\chi}^A ~-~ \frac{1}{a^2} \Delta \chi^A ~+~ 3H \chidot^A~+~\delta^{AB}(U+W)_{,B}~=~ 0,
 \eeq
 where  the inflaton dependence of the interaction potential is as a function of the coherently oscillating 
 background inflaton (here the target space of the decay channel fields has been chosen to be flat).
   The specifics of the oscillations of the components $\phi^I$ depend on the 
 details of the potential and have been discussed in the literature only in the context of separable 
 potentials. This dependence is nontrivial even in general single field monomial inflation \cite{t83}, which 
 explains the focus on separable potentials in the literature. 
 
 In the perturbative theory of reheating \cite{a82etal, afw82, dl82} the effect of particle production is encoded in the Klein-Gordon 
equation by the introduction of a decay term $\G_{\phi^I} \phidot^I$ of the inflaton $\phi^I$ that arises from couplings of 
the inflaton to other fields that provide the decay channels
 $
  \G^I = n^I\si^I v^I
 $
 of the  inflaton components $\phi^I$, leading  to the effective dynamics \cite{st84, kt90}
   \beq
  D_t\phidot^I ~+~( 3H + \G^I) \phidot^I ~+~ G^{IJ}V_{,J} ~=~ 0.
 \eeq
The spacetime expansion is determined by the first Friedmann-Lemaitre equation as
\beq
   H^2 ~=~ \frac{1}{3M_\rmPl^2} \left(\frac{1}{2}G_{IJ} \phidot^I \phidot^J ~+~ V(\phi^I)~+~ \rho_r\right),
  \eeq
  where $M_\rmPl$ is the reduced Planck mass, and the potential dominates the kinetic term for large field values.
 The density of the relativistic decay products, denoted by $\rho_r$, is determined by \cite{st84}
 \beq
  \rhodot_r ~+~ 4H \rho_r ~=~ \sum_I \G^I ( \phidot^I)^2.
  \eeq

\section{Weakly curved trajectories in inflaton target spaces}

The strategy adopted below for analyzing reheating after $j$-inflation is formulated in the present section in a general conceptual way.
The observation is that for some multifield potentials 
 \beq
  V(\phi^I) ~=~ \Lambda^4 F(\phi^I/\mu) 
 \eeq
 described by a dimensionless function $F(\phi^I/\mu)$ that can encode complicated interactions between the multiplet components
 $\phi^I$, it can still happen that  the inflationary trajectories are only weakly curved and hence 
traverse a cross section of the potential that is approximately one-dimensional. Without loss of generality one can characterize 
such a situation in the multifield topography of the potential $V(\phi^J), J=1,...,N$ as a path for which 
 \beq
  \Delta \phi^J ~\ll ~ \Delta \phi^I, ~~~~~\forall J\neq I.
  \eeq
The resulting cross section of the potential at 
 \beq
 V_b(\phi^I) ~=~ V{\Big |}_{\phi^J \cong \rmconst, ~\forall J\neq I}
 \eeq
 leads to a potential that approximates the exact potential in this situation.
 
 With the notation $\phi= \phi^I$  the potential $V(\phi)$ along the inflaton trajectory can be approximated by truncating 
 the  Taylor expansion around its minimum $\phi_0$ at some order. For small $(\phi-\phi_0)$ the lowest order
 is most relevant for the oscillatory behavior of the inflaton and higher orders are suppressed. The potential thus will 
 be of monomial form 
  \beq
   V_b(\phi) ~=~ \Lambda_b^4 \left(\frac{\phi- \phi_0}{\mu}\right)^n
  \eeq
   for some energy scales $\Lambda_b$ and $\mu$. This allows to apply results from the  analysis 
     of monomial inflation based reheating discussions.  For quadratic and quartic inflation 
 the oscillating inflaton $\phi_\rmosc$ can, for example, be described explicitly in terms of trigonometric and Jacobi elliptic 
 functions, respectively \cite{t83, g97etal}.  The standard couplings considered are the triscalar, bi-quadratic,
 and Yukawa couplings 
  \beq
   \cL_\rmint ~=~ - g \phi \chi^2 ~-~ \frac{1}{2}h \phi^2 \chi^2 - h_\psi \phi \opsi \psi,
  \eeq
  with the associated perturbative decay rates $\G_\rmtri$, $\G_{\rm bi.q}$ and $\G_Y$.  These determine the 
  decrease of the amplitude $\cA(t)$ of $\phi_\rmosc$ in addition to the damping that results from the 
  Hubble-Slipher expansion. The decay into radiation $\rho_r$ can then be parametrized as
  \beq
   \rhodot_r~+~ 4H\rho_r ~=~ \G \phidot^2.
  \eeq
    While the bi-quadratic potential has been the focus of much of the preheating literature, it cannot complete 
    reheating, and the triscalar coupling plays an important role toward the end of the process. The fermionic 
    decay channel will be considered further below. In quadratic inflation with 
      mass $m_\phi$ one finds for the triscalar coupling the 
     decay constant  $\G_\rmtri = g^2/8\pi m_\phi$ and the amplitude of the oscillating inflaton decreases as
   \beq
    \cA(t) ~\cong ~ \frac{M_\rmPl e^{-\G_\rmtri t}}{m_\phi t}.
   \eeq
   The result for $\G_\rmtri$ assumes that the computation can be done with perturbative quantum field theory,
  neglecting potential Bose enhancements in the presence of other particles \cite{tb90, dk90, kls94, stb94, kls97}. 
  Since the focus in the following is on  the reheat temperature, such preheating effects will be neglected.
  
  Given the decay rate $\G$ of the inflaton the reheat temperature $T_\rmrh$ is usually computed under the assumption
  of thermalization at the time $t_\G \cong \G^{-1} \cong (3H)^{-1}$. With  the density $\rho_r$ of a 
  distribution of relativistic particles in thermal equilibrium 
    \beq
     \rho_r ~=~ \frac{\pi^2}{30} g_* T^4,
    \eeq
  where $g_* = n_b + (7/8) n_f$ is a count of the number of relativistic degrees of freedom, and evaluating 
  the Friedman-Lemaitre equation
   at $t_\G$
   \beq
    \rho_r ~=~ 3M_\rmPl^2 H^2 ~\cong ~ \frac{1}{3} \G^2 M_\rmPl^2, 
   \eeq
  leads to the reheat temperature  \cite{t83}
     \beq
    T_\rmrh ~=~ \left(\frac{10}{\pi^2 g_*}\right)^{1/4} \sqrt{ \G M_\rmPl}.
   \eeq
  
    The reheat temperature is weakly 
   constrained by nucleosynthesis, which leads to a lower limit of a few MeV. If supersymmetric models are considered,
   there is an upper bound that arises from the gravitino abundance \cite{kl84, kmy06, s08}.
 The set-up just described of an effective reduction of a multifield inflationary dynamics to a one-dimensional reheating 
   epoch is realized in the model of $j$-inflation, to which this is applied below.
      
  Given the reheat temperature the baryon asymmetry can be determined in terms of the parameter $\e$ characterizing 
  the strength of the CP violation. This is usually defined in the context of nucleosynthesis via the photon
   density $n_\g$ as $\eta = n_B/n_\g$,   or in terms of  the entropy density 
   \beq
   s~=~\frac{2\pi^2}{45} g_*T^3 ~=~ \frac{\pi^4 g_*}{45\zeta(3)} n_\g
  \eeq
   as $n_B/s$, leading to  \cite{afw82, st84}
  \beq
   \frac{n_B}{s} ~\cong ~ \frac{3}{4} \e \frac{T_\rmrh}{m_\phi}.
  \eeq
 Here $m_\phi$ is an effective inflaton mass, which might be the Lagrangian mass, or can be given in terms of other 
  fundamental parameters that characterize the inflatonary model.

\section{Reheating after $j$-inflation}

Automorphic inflation is a framework that is based on symmetry considerations, motivated in part by the goal to provide
a unified conceptual context for the shift symmetry as part of discrete symmetry groups that leaves the field theory 
invariant \cite{rs14, rs15}.
The smallest possible groups of interest are congruence subgroups $\G(N)$ at level $N$ of the full modular group
$\rmSL(2,\mathZ)$, leading to the subclass of modular inflation theories \cite{rs14, rs16}.
In modular inflation the potential is determined in terms of a modular function $F$ that is 
constructed from modular forms $\Phi_i$ relative to a congruence subgroup $\G(N)$. 
 The defining congruence relation for a given integer $N$ is not unique and different types of groups $\G(N)$
can be considered. A natural class of potentials is given by 
 \beq
  V_d ~=~ \Lambda^4 |F(\Phi_i)|^{2d}
  \eeq
  for some integer $d$.  
  
  The simplest case is obtained at level $N=1$, in which case the group $\G(1)$ is the full modular group and the symmetry
  imposes the strongest possible constraints on the potential in the modular context.  For this 
  group the set of all modular forms is generated by only two functions, the Eisenstein series of weight four and six. 
  For general weight $w$ the Eisenstein series $E_w$  can be defined in a computationally 
  convenient way in terms of the Bernoulli numbers $B_w$ 
 and the divisor function $\si_w$, given by
  \beq
   \si_w(n) ~=~ \sum_{d|n} d^w,
  \eeq
as
 \beq
  E_w(\tau) ~=~ 1 - \frac{2w}{B_w} \sum_n \si_{w-1}(n) q^n,
  \eeq
 where $q=e^{2\pi i \tau}$. The $E_w$ are modular forms of weight $w$ for $w\geq 4$, but $E_2$ is only a 
  quasimodular form of weight two, characterized by a nonhomogeneous transformation behavior (see below). 
  
  While the potentials of the above type relative to the modular group thus can be written in terms of the modular forms
  $E_4$ and $E_6$,  the  phenomenological analysis of any modular inflation models  also involves 
   the Eisenstein series $E_2$ because  it enters in  the derivatives of modular forms. 
   The space of $\rmSL(2,\mathZ)$-modular forms  and their derivatives is therefore generated by $E_2,E_4$ and $E_6$.
   CMB observables such as the spectral index involve derivatives of the potential, hence of modular forms, and therefore 
   involve $E_2$.
  This raises the issue of the precise modular nature of such observables. It was shown in \cite{rs16} that the 
  nontrivial geometry of the target space provides the necessary structure to  turn the observables  
  into functions whose fundamental building blocks include almost holomorphic terms. Physical observables in 
  modular invariant inflation models are therefore in general not modular, but  almost holomorphic  modular.

  For the full modular group there is a distinguished modular invariant function that is fundamental in the sense that 
  every other modular invariant function of this group can be expressed in terms of this function as a polynomial 
  quotient. This function is the absolute invariant $j(\tau)$ with $\tau$ in the upper halfplane, which therefore provides the 
  basic building block of all functions invariant under $\rmSL(2,\mathZ)$.  Since $E_4$ and $E_6$ are 
  generators of the modular forms, the $j$-function can be written purely in terms of these functions in a 
  computationally convenient way as
  \beq
    j(\phi^I) ~=~ 1728 \frac{E_4^3}{E_4^3 -E_6^2},
  \eeq
  where $\tau = \tau^1+i\tau^2$ is the complex dimensionless inflaton defined as $\tau= \phi/\mu$, with $\mu$ 
  an energy scale. Here the Bernoulli numbers that enter $E_4$ and $E_6$ are given by $B_4=-1/30$ and $B_6=1/42$. 
  A different view of the $j$-function can be obtained in terms of a counting function associated to the harmonic oscillator because
  the denominator in the above definition is related to the Dedekind eta function $\eta(\tau)$ as
   $(E_4^3-E_6^2)/1728 = \eta^{24} = \Delta$, where $\Delta$ is the Ramanujan cusp form of weight twelve.
  
 Given the fundamental nature of the $j$-function, it is natural to consider an inflationary model based on $j$, 
leading to  the potential of $j$-inflation as \cite{rs14,rs16}
 \beq
  V(\phi^I) ~=~ \Lambda^4 |j(\tau^I)|^2,
  \eeq
  where $\Lambda$ is a second energy scale. 
    
  The definition of  modular inflation models is completed by noting that 
  the target space spanned  by the inflaton multiplet is the complex upper half-plane $\cH$. 
  This space has a nontrivial metric geometry $ds^2= G_{IJ} d\tau^I d\tau^J$, determined by the continuous 
  M\"obius group $\rmSL(2,\mathR)$, and  given by  the  Poincar\'e metric
    \beq
     ds^2 ~=~ \frac{(d\tau^1)^2 + (d\tau^2)^2}{(\tau^2)^2}.
  \eeq
  A more detailed discussion of the framework of modular inflation can be found in \cite{rs14, rs16} and a more in-depth 
  discussion of the automorphic generalization to multifield  inflation is given in \cite{rs15}.
  
  In $j$-inflation the scenario described in the previous section is realized in that one of the components 
   of the inflaton multiplet is distinguished because of its larger variation along the orbit. In the case of 
 $j$-inflation it is the first component $\phi^1 = \mu \tau^1$ that varies much more than the imaginary component. 
 This suggests an approximation of  the potential by a monomial fit. As a first approximation it is useful to consider 
 a simple quadratic fit because this allows to apply results known from the simplest chaotic inflation model to the 
 reheating theory of $j$-inflation. By far the most effort in the reheating literature has been focused on this model, 
 with just a few prominent references given by \cite{stb94, kls97}. 
 
The approximation of the potential along the trajectory can, with $\phi = \phi^1 = \mu \tau^1$, be written as
 \beq
  V_j(\phi) ~=~ \frac{\Lambda_j^4}{\mu^2} \left(\phi - \frac{\mu}{2}\right)^2,
 \eeq
  hence degree two chaotic results can be adopted by introducing the effective $j$-inflation mass 
  as
     \beq
    m_j^2~ =~ \frac{2\Lambda_j^4}{\mu^2},
   \eeq
   where $\Lambda_j = \Lambda \a_j$ with $\a_j\cong 59$.
  This mass determines the frequency of the oscillating background inflaton that is usually coupled to its decay channel 
  $\chi$ by either triscalar couplings $g \phi \chi^2$ or bi-quadratic couplings $h\phi^2\chi^2$.
   As noted above, the latter coupling is not 
  sufficient to complete reheating. The decay rate $\G_\rmtri$ in $j$-inflation takes the 
  form
   \beq
    \G_\rmtri^j ~=~ \frac{g^2\mu}{8\sqrt{2}\pi \Lambda^2}
  \eeq
  and therefore the reheat temperature is given by 
   \beq
    T_\rmrh^j ~=~ \frac{1}{2\pi} \left(\frac{5}{4 g_*}\right)^{1/4} \left(\frac{g}{\Lambda_j}\right)  \sqrt{\mu M_\rmPl}
  \eeq
  
 With $T_\rmrh$ the baryon asymmetry after $j$-inflation is given in terms of the fundamental
  parameters of the model as
  \beq
   \frac{n_B}{s} ~= ~ \frac{3}{16 \pi}\left(\frac{5}{g_*}\right)^{1/4} \left( \frac{g\mu}{\Lambda_j^3}\right)  \sqrt{\mu M_\rmPl} ~\e.
  \eeq
  
Only weak constraints are known about the coupling $g$, but the standard suppression of 
quantum corrections of the potential  via $g\ll m_\phi$ is conveniently parametrized as $g = 10^{-\ell}m_j$, where $\ell \geq 1$.
 This leads to estimates of the reheat temperature and the baryon asymmetry in terms of the basic parameters as
 \bea
  T_\rmrh^j &=& \frac{1}{2\pi} \left(\frac{5}{g_*} \right)^{1/4} 10^{-\ell } \Lambda_j \sqrt{\frac{M_\rmPl}{\mu}} \nn \\
  \frac{n_B}{s} &=& \frac{3}{8\pi} \left(\frac{5}{4g_*}\right)^{1/4} \left(\frac{10^{-\ell}}{\Lambda_j}\right) ~\e ~ \sqrt{\mu M_\rmPl}.
  \eea
 Inflationary trajectories with $\mu \cong 10M_\rmPl$ typically lead via the CMB constraints of the scalar amplitude 
\cite{planck15-13, planck15-20} to $\Lambda \cong 10^{-5}M_\rmPl$. Using these ingredients and $g_* \cong 100$ 
we obtain for the reheat temperature after $j$-inflation the estimate
\beq
 T_\rmrh^j ~\cong ~ 10^{13-\ell} \rmGeV  \left(\frac{\Lambda}{10^{13} \rmGeV}\right)
    \left(\frac{M_\rmPl}{\mu}\right)^{1/2},
 \eeq
and for the baryon asymmetry 
  \beq
   \frac{n_B}{s} ~\cong ~ 10^{2-\ell} \e    \left(\frac{10^{13}\rmGeV}{\Lambda}\right)  \left(\frac{\mu}{M_\rmPl}\right)^{1/2} .
  \eeq
  By adjusting the parameters  it is possible to obtain the observed value  $n_B/s \cong 10^{-10}$. 
  
  A similar computation can 
  be done for the Yukawa coupling $h_\psi \phi \opsi \psi$ with decay rate for $m_\psi \ll m_\phi$ as
   \beq
    \G_{\phi \psi} ~=~ \frac{h_\psi^2 m_\phi}{8\pi}.
    \eeq
    The inclusion of this coupling does not raise the reheat temperature because the fermionic reheat temperatures associated
    to $\G_{\phi \psi}$ is suppressed by a factor 
    \beq
    s_p ~\cong ~ \frac{\Lambda_j}{\sqrt{\mu M_\rmPl}}.
    \eeq
    
  The above results for the reheat temperature and the baryon asymmetry show that $j$-inflation can accommodate 
  a low reheating temperature. This makes it natural to consider nonthermal leptogenesis as a mechanism to generate the 
  baryon asymmetry in this framework,  obtained via the inflaton decay into 
  heavy right-handed neutrinos \cite{ls90}. The decay of these Majorana neutrinos into lepton and Higgs fields creates a lepton 
  asymmetry $n_L/s$, which is partially converted into the baryon asymmetry  via sphaleron 
  processes \cite{krs85}.
  This mechanism can avoid abundance problems that one encounters in  baryogenesis models with high reheat temperature.

\section{Conclusion}

The general case of multifield inflation models involves interacting fields, hence  potentials that are not separable 
in the inflaton components, making some aspects of their analysis difficult for an analytical approach. 
A special class of such models is given by theories which exhibit  a simplifying feature in that the topology in field 
space is such that along the inflaton trajectory  the  different components of the inflaton multiplet cover different 
ranges in field space, leading to weakly curved trajectories that admit a quasi one-dimensional treatment.
In this case the potential $V_b$ along  the trajectory depends essentially on a single component of the inflaton multiplet. 

A  systematic class of  highly interacting inflation models is obtained in the context of twofield inflation by the 
framework  of modular inflation.  The model of  $j$-inflation provides an example  of the above phenomenon, which allows to
derive the reheat temperature $T_\rmrh$ and its associated baryon asymmetry $n_B/s$ in terms of the model parameters
and a parameter that tracks the scaling of the coupling parameter. The latter parameter is constrained in order to suppress 
quantum corrections to the potential, leading to an upper bound of the reheating temperature after $j$-inflation of the 
order of $T_\rmrh^j \sim 10^{13-\ell} \rmGeV$.

\vskip .3truein

{\bf Acknowledgement.} \\
It is a pleasure to thank Monika Lynker for discussions. This work was supported by an IUSB research grant.

\end{document}